\documentclass[aps,pra, twocolumn]{revtex4-1}
\usepackage{amsthm}

\usepackage[utf8]{inputenc}
\usepackage[sc,osf]{mathpazo}

\usepackage[T1]{fontenc}
\usepackage{amsfonts}
\usepackage{amssymb}
\usepackage{amsmath}
\usepackage{amsthm}
\usepackage{graphics}
\usepackage{graphicx}
\usepackage{braket}
\usepackage[colorlinks=true,linkcolor=blue,
urlcolor=blue,citecolor=blue]{hyperref}

\begin{document}

%\title{Sequential Measurements  in higher dimension does not give advantage in higher dimension  }
\title{  Robustness of Higher Dimensional Nonlocality  against dual noise\\ and sequential measurements  }

%\title{Separability under global unitary: Bipartite and multipartite cases}

\author{Saptarshi Roy$^{1,2}$, Asmita Kumari$^{1,3}$, Shiladitya Mal$^{1,4,5}$, Aditi Sen (De)$^{1}$}
%\email{x@mail.com}
\affiliation{$^1$ Quantum Information and Computation Group, Harish-Chandra Research Institute, HBNI, Chhatnag Road, Jhunsi, Allahabad 211 019, India}
\affiliation{\(^2\) Quantum Information and Computation Initiative, Department of Computer Science, The University of Hong Kong, Pokfulam Road, Hong Kong.}
\affiliation{$^3$ S. N. Bose National Centre for Basic Sciences, Block JD, Sector III, Salt Lake, Kolkata 700106, India}
%\affiliation{\(^3\) HKU-Oxford Joint Laboratory for Quantum Information and Computation.}

\affiliation{$^4$ Department of Physics and Center for Quantum Frontiers of Research and Technology (QFort), National Cheng Kung University, Tainan 701, Taiwan}
\affiliation{$^5$ Centre for Quantum Science and Technology, Chennai Institute of Technology, Chennai 600069, India}
%\author{Shiladitya Mal}
%%\email{y@mail.com}
%\affiliation{Quantum Information and Computation Group, Harish-Chandra Research Institute, HBNI, Chhatnag Road, Jhunsi, Prayagraj (Allahabad) 211 019, India}
%\affiliation{Department of Physics and Center for Quantum Frontiers of Research and Technology (QFort), National Cheng Kung University, Tainan 701, Taiwan}
%
%\author{Aditi Sen(De)}
%%\email{z@mail.com}
%\affiliation{Quantum Information and Computation Group, Harish-Chandra Research Institute, HBNI, Chhatnag Road, Jhunsi, Prayagraj (Allahabad) 211 019, India}

\begin{abstract}

Robustness in the violation of Collins-Linden-Gisin-Masser-Popescu (CGLMP) inequality is investigated from the dual perspective of noise in measurements as well as in states. To quantify it, we introduce a quantity called the area of nonlocal region which reveals a dimensional advantage. Specifically, we report that with the increase of dimension, the maximally violating states show a greater enhancement in the area of nonlocal region in comparison to the maximally entangled states and the scaling of the increment, in this case, grows faster than visibility. Moreover, we examine the robustness in the sequential violation of CGLMP inequality using weak measurements and find that even for higher dimensions,  two observers showing a simultaneous violation of the CGLMP inequality as obtained for two-qubit states persists. We notice that the complementarity between information gain and disturbance by measurements is manifested by the decrease of the visibility in the first round and the increase of the same in the second round with dimensions.
Furthermore, the amount of white noise that can be added to a maximally entangled state so that it gives two rounds of the violation, decreases with the dimension, while the same does not appreciably change for the maximally violating states.

%Therefore, we get no dimensional advantage both in the maximal number of rounds where violation persists, and the 
%robustness of sustaining multiround violation on mixing white noise to both $\mathcal{M}_E$ and \mathcal{M}_V.

\end{abstract}
\maketitle

\section{Introduction}
\label{sec_intro}
The journey from the Einstein-Podolski-Rosen paradox \cite{epr'35} to Bell theorem \cite{bell'64} via Bohmian mechanics \cite{bohm'52} is a fascinating story that contributed towards our present outlook of a physical theory. It asserts that a satisfactory description of nature cannot assume both the assumptions of locality and reality simultaneously. Jointly these two assumptions, known as local-realism has recently been refuted experimentally by loophole-free Bell test \cite{shalm, hensen, rosen}. Apart from the foundational significance, Bell-Clauser-Horne-Shimony-Holt (CHSH) inequality \cite{chsh'69} enables the device-independent certification of randomness \cite{dirand'10}, secure key distribution \cite{bhk'05, qkd'07, ekert'91}, detection of entanglement \cite{qent'09} etc.
% which is an essential ingredient for many quantum information processing (QIP) tasks like quantum cryptography \cite{ekert'91}, teleportation \cite{tele'93}, dense coding \cite{dens'92}, remote state preparation \cite{akp'00}.

Going beyond the much-studied simplest Bell scenario involving two settings of measurements for two party with two outcomes, denoted by $(2-2-2)$, new insightful and qualitatively different results have been derived which was otherwise impossible if restricted to the simplest scenario. In particular, violation of local realism is manifested more sharply than  $(2-2-2)$-case via Greenberger-Horne-Zeilinger (GHZ) argument \cite{ghz'89}, which requires at least a three-qubit system. With a suitable choice of binary observables, it has been shown that maximal violation of Bell inequality persists for a singlet state of arbitrary spins \cite{gp'92}, turning down the  belief that
systems may loose their quantumness with increasing system size, thereby leading to the decrease in  violation of Bell inequality \cite{gm'82}.  Later, the dimensional advantage in violation of local realism has been established considering a more general choice of observables \cite{kas'00, kas'01, chen'01}, since dichotomic measurements can not exploit the higher dimensional system with full generality. In the bipartite system of arbitrary local dimension, with two choices of non-degenerate measurements, named as $(2-2-d)$-situation, corresponding Bell inequalities have been derived by Collins-Gisin-Linden-Massar-Popescu (CGLMP) \cite{cglmp'02}, which is violated maximally by a nonmaximally entangled state  \cite{acin'02} for the specific choices of observables   \cite{kas'00, kas'01, chen'01}.
% Later it is also proved that non-maximally entangled states violate CGLMP maximally.
 These tight higher dimensional Bell inequalities \cite{masanes'03} exhibit enhancement in visibility with the increase of dimension, thereby showing more robustness against noise \cite{kas'00, cglmp'02} Here visibility  refers to the noise strength upto which a pure state mixed with white noise exhibits violation. 
 %It is also important to mention here that higher dimensional bipartite systems also turn out to be useful in several quantum information processing tasks ranging from quantum key distribution, quantum dense coding \cite{dens'92} \textcolor{red}{to quantum teleportation \cite{tele'93}.}
  It has been shown explicitly that performances of many quantum information processing tasks get enhanced by considering higher dimensional systems. Specifically, compared to qubits, 
 %higher dimensional systems 
 they exhibit greater robustness to noise \cite{cglmp'02}, stronger security in device-dependent quantum key distribution (QKD) \cite{ddqkd'02} and device-independent extraction of random bits \cite{dirand'22}. However,  dimensional advantage is not always straightforward as noted in a recent work \cite{diqkd'24} in the context of device-independent QKD -- the estimated lower bound on the secure key rate does not improve with increasing the dimension whereas the upper bound on the key rate exhibits the opposite trend. 
 % to computational complexity \cite{dim1,dim2, ddc, dim3, dim4, dimHardy,  dim5, dim6, dim7, dim8, ddc, photonexp, comcomplex, cryptoadv, Sapi20}. 

In another direction, the conventional Bell scenario has been extended where half of a bipartite system is possessed by a single observer, called Alice, while the other half is possessed by a series of observers, named as Bobs, who can measure sequentially \cite{silva'15}. In this new scheme of the Bell test, it has been shown that no more than two observers can violate Bell-CHSH inequality if the series of observers measure independently \cite{silva'15, malchsh}. Such a sequential scenario has also been tested experimentally  \cite{ex1, ex2} and is further extended in several situations which include detecting steerable correlation \cite{malsteer, she'19}, witnessing entanglement \cite{malew, malmdiew}, testing Bell inequalities other than CHSH \cite{malbi}, identifying genuine entanglement \cite{gent}, and preparation contextuality \cite{contex'19}. An interesting twist in this situation is that with the slight modification to the independent and unbiased measurement scheme, an unbounded sequence of observers can be found who can certify non-classical correlation with the single observer in another side \cite{silva'15, colbeck'20}. Recently, some interesting applications of the sequential scheme like self testing unsharp measurement \cite{selftest'19, unsharp'20}, reusing teleportation channel \cite{wktele}, generating randomness \cite{rand'17} have been proposed, thereby showing its potentiality in quantum technologies. 
An interesting observation from the above studies is that if one restricts to a particular measurement scheme, i.e., independent and unbiased measurement by the series of observers \cite{malchsh}, it is not straightforward to predict the number of sequential observers that would show non-local correlations. As the number of sequential violations depends on the initial strength of the correlation, detection and measurement process in a non-trivial way,  it is not well characterized yet.
% the number of successful detection of nonclassical correlation depends on the strength of the underlying correlation, detection and measurement processes in an intricate way which is not well understood yet. 
The number is finite and dictated by the trade-off between the disturbance and information gained by measurements. For example, it was found that for maximally entangled two-qubit initially shared state,  at most twelve Bobs can detect entanglement with a single Alice employing measurement settings pertinent to the optimal witness operator \cite{malew} while two Bobs sharing the same state with Alice can violate CHSH inequality \cite{silva'15, malchsh} 
%with maximally entangled two-qubit state and 
based on optimal measurement settings required for Bell-violation. In the sequential measurement, partial information is extracted which is sufficient for the detection scheme and at the same time,  some residual correlation remains for other rounds which gradually diminishes with a longer sequence of Bobs.
% It indicates not only that weaker correlation survives more
It also reveals that witnessing entanglement possibly disturbs the state less compared to the situation when the Bell-CHSH test is performed, thereby admitting more robustness of the former scheme against noise. Similarly, measurement-device-independent entanglement witness \cite{mdiew'13} turns out to be more suitable in the sequential situation than that of the standard entanglement witness \cite{ew} as shown through the increased number of Bobs \cite{malmdiew}. 

In the present work, we first investigate the robustness of CGLMP inequality by going beyond the visibility measure of `nonlocality' \cite{kas'00, cglmp'02}. Specifically, in addition to white noise in the state, we consider noisy measurement (which we call as weak/unsharp measurement) on the maximally entangled state ($\mathcal{M}_E$) as well as on the maximally CGLMP violating states ($\mathcal{M}_V$). Such a consideration of dual noise leads to a measure of robustness, dubbed as the `area of nonlocal region' (where nonlocality means the violation of CGLMP inequality),  which scales with dimension more sharply than the visibility one. The introduction of noise to the measurement enables the possibility of sequential violation of CGLMP inequality.  In particular, we find that the violation by two  Bobs'  persists even with the increase of dimension, as found in the two-qubit case with CHSH inequality. In this respect, the pertinent question is how the robustness of CGLMP is reflected in the sequential scenario.
%when the shared state is maximally as well as  nonmaximally entangled two-qudit state.
 It was noticed that in the context of a violation of CGLMP inequality, the visibility decreases with the increase of dimension \cite{kas'00, cglmp'02}. However, we observe that if we demand the violation of CGLMP inequality in two rounds of a sequential scheme, the required visibility increases with the dimension for maximally entangled states while surprisingly, it remains constant for maximally violating states.
%to obtain double violation of CGLMP, the required minimum visibility increases with dimension, 
%thereby showing a contrasting behavior between a normal Bell test and a sequential  protocol.  
It also demonstrates that the sequential scenario can reveal a kind of robustness which is qualitatively different from the visibility and `area of nonlocal region' obtained for a single round. It is due to the trade-off present in the disturbance by the weak measurements and the information gain via measurements in a sequential scheme. 
%We analyse CGLMP upto dimension ten for both maximally entangled as well as  non maximally entangled states.

The paper is organised in the following way. In Sec. \ref{sec_Bell ineq}, we briefly discuss the prerequisite of the present work. In Sec. \ref{sec:robustregion}, the robustness of CGLMP is discussed with a new measure introducing dual noise. For higher dimensional pure states, CGLMP inequality is used to certify entanglement sequentially in Sec. \ref{sec:seq:pure} and a similar study is carried out for noisy mixed states in Sec. \ref{sec:seqnoise}.   We conclude in Sec. \ref{sec_disc} with a brief discussion.

%{\color{red}Higher dimensional system has also been shown to be advantageous in ...such as implementation of Toffoli gate require lesser gate if three level system is involved. I add few lines later}

\section{Prerequisites: Bell inequalities in higher dimensions and Sequential measurement scheme}
\label{sec_Bell ineq}

Before presenting our results, let us briefly discuss the CGLMP inequality and sequential scenario of the Bell test.

\subsection{CGLMP inequality}
 Let Alice and Bob be two observers allowed to perform two $d$ outcome measurements. If $A_1$ and $A_2$ are measurement settings of Alice while $B_1$ and $B_2$ are of Bob, which can take values from $ [0, d-1]$, i.e., $A_{1(2)}, B_{1(2)} = 0,1, \ldots, d-1$.  The CGLMP inequality reads as \cite{cglmp'02}
 \begin{eqnarray}
\label{cgl}
I_d &=&  \sum^{\left\lfloor \frac{d}{2}\right\rfloor -1}_{k = 0} \Big( 1 - \frac{2k}{d-1} \Big) [f(k)-f(-k-1)]  \leq 2,
\end{eqnarray}
where
\begin{eqnarray}
f(k) &=& P(A_1 = B_1 +k) + P(B_1 = A_2+k+1) \\ \nonumber &+& P(A_2 = B_2+k)+ P(B_2 = A_1+k). 
\end{eqnarray}
The probabilities of the outcomes of Alice's measurement, $A_a$ and Bob's measurement, $B_b$, ($a,b = 1, 2$) in $f(k)$ differ by $k$ mod $d$ and can be written as
\begin{eqnarray}
\nonumber
P(A_1 = B_1 +k) = \sum^{d -1}_{j = 0} P(A_a =j, B_b = j+ k \ \ mod \  \ d).
\end{eqnarray}
The strongest violation of CGLMP inequality is obtained for a maximally entangled state and a particular class of non-maximally entangled states if the 
%measurement settings on Alice and Bob are chosen to be 
measurements performed by Alice and Bob are in the basis $\{|k \rangle_{A_a}\}$ and $\{ |l \rangle_{B_b} \}$ with
\begin{eqnarray}
\label{ma}
|k \rangle_{A_a} = \frac{1}{\sqrt{d}}\sum^{d-1}_{j=0}\exp(i \frac{2 \pi}{d}j(k + \alpha_a))|j \rangle_{A},
\end{eqnarray}
and
\begin{eqnarray}
\label{mb}
|l \rangle_{B_b} = \frac{1}{\sqrt{d}}\sum^{d-1}_{j=0}\exp(i \frac{2 \pi}{d}j(-l + \beta_b))|j \rangle_{B}, 
\end{eqnarray}
where 
\begin{eqnarray}
\label{v1}
 \alpha_1= 0, \ \ \ \alpha_{2} = 1/2,\ \ \  \beta_{1} = 1/4 \ \ \  and  \ \ \   \beta_{2} = -1/4.
\end{eqnarray}
  The special thing about the above inequality is that its quantum violation increases with dimension $d$. %In this paper, the issue of robustness in the CGLMP inequality is addressed  in two ways  -- one is by studying the trade-off in noise given in states and in measurements while the other one  is by considering the sequential measurements which we will now briefly describe.   

%  It was shown that if one considers
% \begin{eqnarray}
% \rho = p |\psi\rangle \langle \psi| + I/d,
% \end{eqnarray}
% where \(|\psi\rangle\) is maximally entangled state, then the visibility, \(p\), requires to obtain violation of CGLMP decreases with the dimension, \(d\), thereby showing the robustness of this inequality with dimension \cite{cglmp'02, kas'00}. 
% In order to know the importance of increasing violation of CGLMP inequality with increasing dimension, in this paper we will study in two ways. Since robustness of both the state and measurement play an important role for the violation of any Bell-type inequality, we perform the detailed study of the robustness for the violation of  CGLMP inequality both for maximally and non-maximally entangled state up to $d=10$. Then we discuss the sharing of non-locality among several observers using CGLMP inequality up to $d=10$. Since for sharing non-locality, each observer measure sequentially on her part,   we will briefly describe the sequential measurement scenario, before proceeding further.

\subsection{Sequential measurement scenario}

The sequential measurement scenario considers an entangled state of two $d$-dimensional systems shared in such a way that half of the system is in possession of the observer (say, Alice) and the other half is with several observers (say, $n$ Bobs, referred as Bob$_1$, Bob$_2$, Bob$_3$, \(\dots\), Bob$_n$). The task of Bob$_1$ is to pass the system to Bob$_2$ after performing an unsharp measurement on his part. Similarly, Bob$_2$ passes the system to Bob$_3$  after the measurement
%after performing the unsharp measurement on the state sent by Bob$_1$ 
and so on. In other words, several Bobs measure their part sequentially, hence the name of the sequential measurement scheme. Note that the measurement of each Bob is independent and all the measurement settings of each Bob are equally probable. In the scenario described above, if the measurement statistics between Alice and any Bob, say, Bob$_k$ \(Bob_k\) (\(k>1\))  exhibits a violation of CGLMP, then we call it as 'sharing of nonlocality' between them in the sequential scenario.

 To know the number of Bob sharing nonlocality of a shared entangled state (say, $\rho$) between Alice and $n$ Bobs, we have to assume that the measurement of Alice and  Bob$_n$ is sharp (i.e., they perform projection measurements on their parts). In contrast, $1, \ldots, n-1$ Bobs perform unsharp measurements represented by positive-operator valued measurements (POVMs). 
If measurement settings at Alice are denoted by $\{|k \rangle_{A} \langle k |\}$  and the measurement settings of Bob$_m$ is represented by 
\begin{eqnarray}
\label{pv}
E^{l}_{B_{m}} = \lambda_m |l \rangle_{B} \langle l| + \frac{1-\lambda_m}{d} \mathbb{I}_d,
\end{eqnarray}
where $k, l = 0,1,2,d-1$, $m = 1, 2, 3, 4.....n-1$, $\lambda_m$ ($0 < \lambda_m \leq 1$) is the sharpness parameter of Bob$_m$ and $\mathbb{I}_d$ is the $d$-dimensional identity matrix. The  state after the measurements of $(m-1)$th Bob and without any measurement at Alice's end transforms as
\begin{eqnarray}
\label{state}
\rho_{m} = \frac{1}{d}\sum^{d-1}_{l = 0}(\mathbb{I}_d \otimes \sqrt{E^{l}_{B_{m-1}}}) \rho_{m-1} (\mathbb{I}_d \otimes \sqrt{E^{l}_{B_{m-1}}}),
\end{eqnarray}
where \(\rho_{m-1}\) is the state before the unsharp measurement performed by Bob$_{m-1}$.
%\textcolor{red}{I think, the above equation is not correct. CHANGE IT}
We will use the post measured state $\rho_{m}$ and POVM in Eqs. (\ref{pv}) and  (\ref{state}) respectively, when we certify nonlocality via CGLMP inequality in this scenario.

\section{Robustness in CGLMP violation: Area of nonlocal region }
\label{sec:robustregion}

The study of the violation of Bell-type inequalities is a major endeavour in studies of nonlocality. Another important aspect is the investigation of robustness in the obtained violation. Typical studies of robustness consist of the addition of noise to the state and tracking the response of violation due to the amount of noise added to the state. However, for the violation of Bell-type inequalities, measurements play as crucial a role as states. Therefore, robustness analysis should also be carried out when noise is added to the measurements as well. 

We perform a general robustness analysis when both the state as well as the measurements are simultaneously noisy. In particular, we explore the role of the dimension of the bipartite state whose nonlocal characteristics in terms of violation of CGLMP inequality in Eq. (\ref{cgl}) are under investigation.    Before that, we briefly discuss the scenario when white noise is mixed with the state, given by
\begin{eqnarray}
\rho = p|\psi\rangle\langle \psi| + \frac{1-p}{d^2}\mathbb{I}_{d^2},
\label{eq:noisystate}
\end{eqnarray}
where $|\psi\rangle$ is a bipartite pure state with each party of dimension $d$, and  $\mathbb{I}_{d^2}$ is the $d \otimes d$ maximally mixed state (white noise). It was observed \cite{cglmp'02} that when $|\psi\rangle$ is a maximally entangled state in  $d \otimes d$,  given by \(|\psi_{\mathcal{M}_E}^d\rangle = \frac{1}{\sqrt{d}}\sum_i | ii \rangle\), the 
robustness to noise which can be called as visibility, measured as $p$, increases with the increase in $d$. This is in the sense that the 
maximal white noise that can be added to $|\psi_{\mathcal{M}_E}^d\rangle$ such that the resultant mixed state $\rho$ violates the CGLMP inequality which  increases with dimension $d$. For a given $d$, this maximal amount of white noise is denoted by $1 - p_{\min}$. In other words, for $|\psi_{\mathcal{M}_E}^d\rangle$, as $p_{\min}$ decreases, $1 - p_{\min}$ increases with $d$. Such dimensional advantage of robustness is enhanced when instead of a $\mathcal{M}_E$, $|\psi\rangle$ is chosen to be the non-maximally entangled state which violates CGLMP inequality maximally \cite{acin'02}. We call such maximally violating states as $\mathcal{M}_V$. The exact form of $\mathcal{M}_V$ up to $d = 10$  can be found in Ref. \cite{Fonseca'18}.  The $\mathcal{M}_V$ offer a greater robustness with $d$ in comparison to $\mathcal{M}_E$. For the convenience of readers, we present the exact form of the maximally violating states in Appendix \ref{app:1}.

\begin{figure*}[ht]
\includegraphics[width=\linewidth]{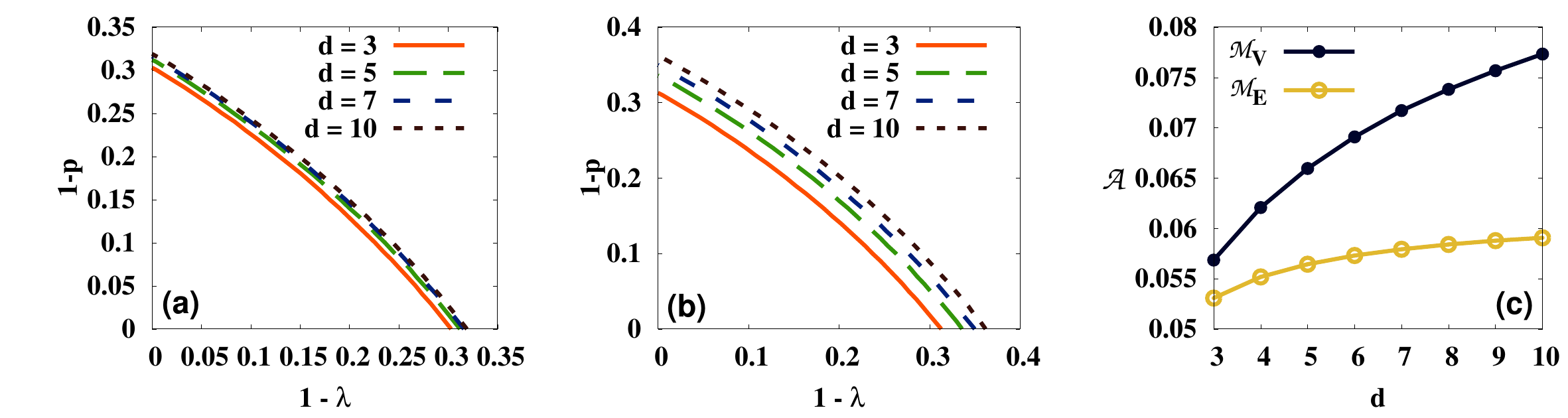}
\caption{(a) For a fixed $d$, each point in the curve just crosses the local realist value of $2$ i.e., when the value of \(I_d\) in Eq. (\ref{cgl}) is just above \(2\) by taking maximally entangled state ($\mathcal{M}_E$) in the $(1-\lambda, 1-p)$-plane. Different $d$ values are considered. (b) Similar plot when the shared state is maximally CGLMP  violating state ($\mathcal{M}_V$). 
(c) $\mathcal{A}$  (ordinate) defined in Eq. (\ref{eq:ANR}) vs. $d$ (abscissa) for $\mathcal{M}_E$ and $\mathcal{M}_V$. 
%For a given $d$, all configurations under the curve correspond to violations of CGLMP inequality.
}
\label{fig:robust1}
\end{figure*}

We now consider the opposite situation where the shared state is noise free and the measurements the measurements are taken to be noisy.
The effect operators for the noisy measurements are described by POVMs given in Eq. (\ref{pv}). Note that here we are interested in the first round violation $(m = 1)$, see Eq. \eqref{pv}.
Considering $|\psi\rangle$ to be noiseless $\mathcal{M}_E$ and $\mathcal{M}_V$, the amount by which measurements can be made noisy preserving violation of the CGLMP inequality is denoted by $1 - \lambda_{\min}$, which also increases with increasing  $d$. We observe an exactly similar dimensional dependence with noise in the measurements denoted by $\lambda$, as obtained in the case of noisy states
%. This is because of the equivalence of 
adding white noise to the state or measurements is equivalent. Mathematically, for any pure state $|\psi\rangle$,
\begin{eqnarray}
I_d^Q(p = 1, \lambda = x) = I_d^Q(p = x, \lambda = 1).
\label{eq:syymetry}
\end{eqnarray}
Here superscript $Q$ in  Eq. \eqref{eq:syymetry} implies that the probabilities associated with the CGLMP expression are calculated for the quantum states and measurements performed by Alice and Bob, stated earlier. For brevity, we do not henceforth use the superscript $Q$ as we always work with probabilities generated by quantum states and measurements.
Things become more interesting and involved when both the state and measurements suffer from noise simultaneously, which we will discuss in the next subsection.

\subsection{Complimentarity of Robustness}
\label{sec:robust1}

We are now going to study the robustness obtained from the violation of CGLMP inequality by considering both the state and the measurements noisy. We again start with a $d$-dimensional maximally entangled state, $|\psi_{\mathcal{M}_E}^d \rangle$ as well as $\mathcal{M}_V$, \(|\psi_{\mathcal{M}_V}\rangle\) (for two-qutrit state, $\mathcal{M}_V$ is of the form, $\frac{1}{\sqrt{t}}(|00\rangle +\gamma |11\rangle +|22\rangle)$, where $t=2+\gamma^2$ and $\gamma = 0.7923$ \cite{acin'02, Fonseca'18}). In this general framework, $p_{\min}$ is a function of the noise in the measurements, which we denote as $p_{\min}(\lambda)$, and naturally $\lambda_{\min}$, in turn, becomes a function of the noise added to the state, which is referred to as $\lambda_{\min}(p)$. For convenience, we drop the $\min$ and functional labels,  thereby indicating $1-p_{\min}(\lambda)$ and $1-\lambda_{\min}(p)$ as $1-p$ and $1-\lambda$ respectively. We investigate the dual version of robustness by tracking the locus of all the points in the $(1-\lambda, 1-p)$-plane that just crosses the local realist value of $2$ by considering $\mathcal{M}_E$ and $\mathcal{M}_V$, see Figs. \ref{fig:robust1} (a) and (b). Note that all noise configurations that fall below the curve lead to the violation of the CGLMP inequality. Therefore, in the $(1-\lambda, 1 - p)$-plane, the ratio of the areas under the curve can be considered to be a measure of robustness when both the state and the measurement are affected by noise. Motivated by this observation, we introduce a generalized robustness measure as the area under this curve, which we call the ``area of nonlocal region" ($\mathcal{A}$).
Mathematically, the area of nonlocal region ($\mathcal{A}$) in the noise plane can be defined as
\begin{eqnarray}
\mathcal{A} = \int_0^{1-p_{\min}(\lambda = 1)} \big(1-\lambda_{min}(1-p)\big) ~dp.
\label{eq:ANR}
\end{eqnarray}
We then compute $\mathcal{A}$ values for both $\mathcal{M}_E$ and $\mathcal{M}_V$, and make a comparative analysis of their respective scalings with $d$, see Figs. \ref{fig:robust1} (a), (b), and (c). The \(\mathcal{A}\) values for the $\mathcal{M}_E$ and $\mathcal{M}_V$ are listed in Table \ref{table:ANR}. Our findings are listed below:
\begin{enumerate}
\item The $\mathcal{A}$ values for $\mathcal{M}_V$ are strictly greater than those obtained for $\mathcal{M}_E$. Furthermore, the gap of $\mathcal{A}$ values for the $\mathcal{M}_V$ and the  $\mathcal{M}_E$ grows with $d$ as clearly discernible from Table \ref{table:ANR} and Fig. \ref{fig:robust1} (a) - (c).

%\item With the increase of $d$, the ANR curve changes mu

\item The $\mathcal{A}$ scales much faster with $d$ for the $\mathcal{M}_V$ in comparison to the $\mathcal{M}_E$, see Fig. \ref{fig:robust1} (c).

\item The gap in the growth between $\mathcal{M}_V$ and $\mathcal{M}_E$ in the case of $\mathcal{A}$ grows much faster than that of the visibility. 
\end{enumerate}

From the fact that the value of the CGLMP expression is same when either the state or the measurement becomes noisy with the same amount of white noise, $I_d(p = 1, \lambda = x) = I_d(p = x, \lambda = 1)$ as in Eq. \eqref{eq:syymetry}, it seems reasonable to assume that robustness can be completely characterized by looking at either $p$  or $\lambda$. However, this symmetry does not hold in the general case where both noises are non-vanishing, i.e., $I_d(p = x, \lambda = y) \neq I_d(p = y, \lambda = x)$. It is reflected by the fact that the curves in Figs. \ref{fig:robust1} (a) and (b) are non linear. Therefore, in the general case, the CGLMP nonlocality is both a function of the unsharp parameter and the amount of noise in the state in a non trivial way. This leads us to introduce $\mathcal{A}$ as a single-letter formula to analyze robustness that incorporates the effects of both.

\begin{figure}[ht]
\includegraphics[width=\linewidth]{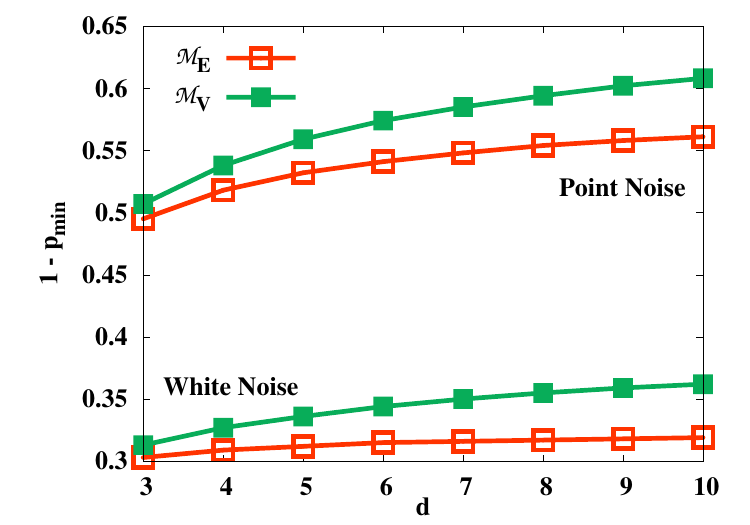}
\caption{Comparison of robustness to point and white noise of the maximally entangled states $\mathcal{M}_E$ and the maximally violating ones $\mathcal{M}_V$ for $d = 3$. Both $\mathcal{M}_E$ and $\mathcal{M}_V$ are more robust to point noise as indicated by a higher value of $1 - p_{min}$.}
\label{fig:n2}
\end{figure}

Furthermore, note that the white noise paradigm can be motivated by interpreting the noise is arising from a depolarizing noisy channel that is independent of the source state and can be interpreted as a systematic noise involved in the setup. This, as we understand, gives a fair platform for comparing the performance of various states with respect to their robustness against noise. For coloured noise, there is no unique way to ascribe the nonuniformity of the noise, and it can alter the non-local properties of various states in completely different ways. It would make our comparative studies hard. This is why we focus on the systematic white noise in both cases. 
In the general case, one intends to consider an arbitrary separable state as noise. Such noise would depend on the initial shared entangled state. To create a meaningful platform of comparison for various initial shared states ($\mathcal{M}_E$, $\mathcal{M}_V$ etc.), for each initial state, one has to perform an optimization over the entire set of separable states pinning down on the least disturbing noise for each initial state. This in general is a very hard problem.      
Nevertheless, we take a particular coloured noise scheme that might be worthwhile to consider, which we call the point noise, where any state $|\psi\rangle$ under consideration is made noisy by
\begin{eqnarray}
\rho = p |\psi\rangle\langle\psi| + (1-p)|k = 0\rangle\langle k = 0|\otimes|l = 0\rangle\langle l = 0|. \nonumber \\
\end{eqnarray}
For the expressions of $|k (l) = 0\rangle$, see Eqs. \eqref{ma} and \eqref{mb}. Compared to white noise (mixing maximally mixed state), both $\mathcal{M}_E$ and $\mathcal{M}_V$ show enhanced robustness features with respect to the point noise which can be easily observed from Fig. \ref{fig:n2} for $d = 3$. However, the scaling of $1 - p_{min}$  with $d$ remain qualitatively similar with the hierarchies between $\mathcal{M}_V$ and $\mathcal{M}_E$ being preserved. For $d = 3$ case, see Fig. \ref{fig:n2}.

\begin{table}[]
\begin{center}
\begin{tabular}
 {|p{1.5cm}|p{1.5cm}|p{1.5cm}|p{1.5cm}|}  
%\hline
%\multicolumn{4}{|c|}{Optimal quantum value of CGLMP inequality} \\
\hline
\centering $d$ & \centering $\mathcal{A}$ ($\mathcal{M}_E$) & \centering $\mathcal{A}$ ($\mathcal{M}_V$) &  \hspace{0.3 cm} Diff. \\
\hline
\centering $3$ & \centering $0.05307$ & \centering $0.05685$ & \hspace{0.25 cm}  $7.14 \%$  \\
\centering $4$ & \centering $0.05517$ & \centering $0.06207$ & \hspace{0.25 cm}  $12.51 \%$  \\
\centering $5$ & \centering $0.05644$ & \centering $0.06595$ & \hspace{0.25 cm} $16.85 \%$  \\
\centering $6$ & \centering $0.0573$ & \centering $0.06909$ & \hspace{0.25 cm} $23.81 \%$  \\
\centering $7$ & \centering $0.05792$ & \centering $0.07171$ &  \hspace{0.25 cm} $27.45 \%$ \\
\centering $8$ & \centering $0.0584$ & \centering $0.07382$ & \hspace{0.25 cm}  $26.40 \%$  \\
\centering $9$ & \centering $0.05878$ & \centering $0.07567$ & \hspace{0.25 cm}  $28.73 \%$  \\
\centering $10$ & \centering $0.05906$ & \centering $0.07733$ & \hspace{0.25 cm}  $30.93 \%$  \\
\hline
\end{tabular}
\end{center}
\caption{The $\mathcal{A}$ values for maximally entangled  ($\mathcal{M}_E$) as well as maximally violating states ($\mathcal{M}_V$), and their percentage differences (labelled as Diff.) from $d=3$ to $d=10$ are given in different columns. The difference grows with $d$ since the $\mathcal{A}$ for $\mathcal{M}_V$ scales much faster with increase in $d$ than that of $\mathcal{M}_E$.}
 \label{table:ANR}
\end{table}

Typically, noise in the system has an adverse effect on the system in the form of lowering the visibility.  As shown in this section, the bane can turn out to be boon in disguise if we look at the situation from a different point of view. In the context of sequential measurements, the ``white noise" in the measurement actually constitutes a POVM strategy which allows multiple Bobs to share nonlocality, thereby manifesting the robustness from a different perspective, as will be shown in the succeeding section. 

%. We would detail this issue in the next section. 

\section{Sharing of nonlocality in higher dimension}
\label{sec:seq:pure}

%Since for $d=2$ CGLMP inequality reduces to CHSH inequality and sharing of non-locality using CHSH is well known results.
In the sharing scenario considered in this section, we deal with the maximally entangled and maximally violating states shared by Alice and Bob$_1$ in arbitrary dimension.    
We will start our discussion from $d=3$ and a detailed analysis is presented for $\mathcal{M}_E$ in $d=3$ to $d=5$. We then repeat the investigation for the maximally violating states.

After substituting $d = 3$ in Eq. (\ref{cgl}), the CGLMP inequality reads as
\begin{eqnarray}
\nonumber
\label{cg3}
I_3 &=& P(A_1 = B_1) + P(B_1 = A_2+1) +P(A_2 = B_2) \\  \nonumber &+& P(B_2 = A_1)- [P(A_1 = B_1-1) + P(B_1 = A_2) \\   &+&P(A_2 = B_2-1)+P(B_2 = A_1-1)] \leq 2.
\end{eqnarray}
%If the state shared between Alice and several Bobs is 
If the shared state is the two-qutrit $\mathcal{M}_E$,  given by
\begin{eqnarray}
\label{p}
|\psi_{\mathcal{M}_E}^3 \rangle =  \frac{1}{\sqrt{3}}(|00 \rangle +  |11 \rangle +  |22 \rangle),
\end{eqnarray}
 by performing POVM at Bob$_1$'s side, and by considering the measurement settings  for  CGLMP test given in Eqs. (\ref{ma}), (\ref{mb}) and (\ref{v1}) for  Alice and Bob$_1$,  the quantum expression of CGLMP inequality, $I_3$ (Eq.(\ref{cg3})) for  Alice-Bob$_1$-pair reduces to 
\begin{eqnarray}
\label{b1m}
 I^1_3 = \frac{4}{9}(3+ 2\sqrt{3}) \lambda_1,
\end{eqnarray}
where the superscript, "\(1\)" represents the number of rounds in the sequential scenario. 
Hence,  the nonlocality can be demonstrated by showing the violation of CGLMP inequality between Alice and Bob$_1$ if $\lambda_1 > 2/(\frac{4}{9}(3+ 2\sqrt{3})) = 0.69615 $ while the optimal quantum value  for Alice and Bob$_1$ is $2.87293 $ obtained at $\lambda_1 = 1$. In a similar fashion, we can find the quantum expressions for  Alice-Bob$_2$ and Alice-Bob$_3$-pairs are respectively  
\begin{eqnarray}
\nonumber
  I^2_3 &=& \frac{4  \lambda_2}{81} \bigg[-2 \left(\sqrt{3}+3\right) \lambda_1+12 \sqrt{1-\lambda_1} \sqrt{2 \lambda_1+1}\\ &+&4 \sqrt{2 \lambda_1+1} \sqrt{3-3 \lambda_1}+14, \sqrt{3}+15\bigg],
\end{eqnarray}
and
\begin{eqnarray}
\nonumber
  I^3_3 &=& \frac{4 \lambda_3}{729} \bigg[4 (\sqrt{3}+6)  \left(2 \sqrt{1-\lambda_2} \sqrt{2 \lambda_2+1}-\lambda_2\right)\\ \nonumber &\times&\sqrt{1-\lambda_1} \sqrt{2 \lambda_1+1}-2 \lambda_1 \big(7 \sqrt{3}+15-\left(\sqrt{3}+6\right) \lambda_2\\ \nonumber &+& 2 (\sqrt{3}+6) \sqrt{1-\lambda_2} \sqrt{2 \lambda_2+1}\big)-2 \left(7 \sqrt{3}+15\right) \lambda_2\\ \nonumber &+&4\left(7 \sqrt{3}+15\right)( \sqrt{1-\lambda_1} \sqrt{2 \lambda_1+1}\\&+&\sqrt{1-\lambda_2} \sqrt{2 \lambda_2+1}+75+98 \sqrt{3})\bigg].
\end{eqnarray}
%From Eq.(\ref{b1m}), we can say that The optimal quantum value for Alice and Bob$_1$ is $2.87293 $ obtained at $\lambda_1 = 1$. 
Considering the situation of minimum violation of $ I^1_3$ by Alice and Bob$_1$, the quantum expression of $I^2_3$ reduces to be $2.40856 \lambda_2$. In this case, the violation of CGLMP inequality for Alice and Bob$_2$ is possible if  $\lambda_2 > 0.830372 $ while the optimal quantum value is $2.40856 $ with $\lambda_2 = 1$.
Substituting the conditions for \(\lambda_1\) and \(\lambda_2\), we get that two Bobs surely violate CGLMP inequality. Let us now check whether the third Bob, Bob$_3$ can also violate CGLMP inequality or not. In this case, the optimal quantum value of $I^3_3$ turns out to be $1.83798 <  2 $
% In case of Bob$_{3}$, even 
by taking minimum violation condition for Bob$_{2}$ and Bob$_{3}$. Since optimal quantum value of $I^3_3$ is strictly less than $2$, we can claim that only two Bobs,  Bob$_{1}$ and Bob$_{2}$, can exhibit nonlocality with Alice by using CGLMP inequality for $d = 3$. Notice here that only two Bobs can violate CHSH inequality with Alice if they initially share a two-qubit maximally entangled state \cite{silva'15}.

Let us now move to $d = 4$ and $d=5$. In this case, \(I_d\) in Eq. (\ref{cgl}) reduces to
% in Eq.(\ref{cgl}), CGLMP inequality for four dimensional system is obtained as
\begin{eqnarray}
I_4 &=& P(A_1 = B_1) + P(B_1 = A_2+1) +P(A_2 = B_2) \\ \nonumber &+& P(B_2 = A_1)  - [P(A_1 = B_1-1) + P(B_1 = A_2) \\ \nonumber &+&P(A_2 = B_2-1)+P(B_2 = A_1-1)] \\ \nonumber &+&  \frac{1}{3}\big(P(A_1 = B_1 + 1) +P(B_1 = A_2+2)\\ \nonumber &+& P(A_2 = B_2 + 1) +P(B_2 = A_1 + 1 ) \\ \nonumber &-& [P(A_1 = B_1-2) P(B_1 = A_2 - 1)\\ \nonumber &+&  P(A_2 = B_2-2) +P(B_2 = A_1-2)]\big)\leq 2.
\end{eqnarray}
%Let us assume that the initial state is maximally entangled given by
%\begin{eqnarray}
%\label{p4}
By following a similar prescription, for a maximally entangled state, 
\(|\psi_{\mathcal{M}_E}^4 \rangle =  \frac{1}{2}(|00 \rangle +  |11 \rangle +  |22 \rangle +  |33 \rangle)
\),
%end{eqnarray}
%and for the measurement settings of Alice and Bobs  in Eqs. (\ref{ma}), (\ref{mb}) and (\ref{v1}),
%Then following the similar procedure of $d=3$, 
we find that Bob$_1$ starts sharing nonlocality with Alice through the violation of CGLMP when  $\lambda_1 > 0.690551$ and  \(\max I_4^1 = 2.89624\) for $\lambda_1 = 1$. Again, %by considering the situation of minimum violation of $ I^1_4$ by Alice and Bob$_1$, we found that
if we restrict the situation such that Alice-Bob$_1$ duo just shows violation, 
%The violation of CGLMP inequality $ I^1_4$ for 
Alice and Bob$_2$ violate CGLMP when $\lambda_2 >  0.834603$ and in the second round, the maximal quantum value is reduced which is $ 2.39635$  ($\lambda_2 = 1$). 
%In case of Bob$_{3}$  
By taking the minimum violation condition of sharpness parameter for Bob$_{2}$ and Bob$_{3}$, the optimal quantum value of $I^3_3$, given in Table \ref{tab:mes},  again tuns out to be less than $2$. 
%Thus similarly as for $d=3$ only two Bobs,  Bob$_{1}$ and Bob$_{2}$ can share non-locality using CGLMP inequality $I_4$ for $d = 4$.
%Further, following the same procedure of $d = 3$ and $d=4$ 
For $d = 5$,
% with CGLMP inequality given by
\begin{eqnarray}
I_5 &=& P(A_1 = B_1) + P(B_1 = A_2+1) +P(A_2 = B_2) \\ \nonumber &+& P(B_2 = A_1)  - [P(A_1 = B_1-1) + P(B_1 = A_2) \\ \nonumber &+&P(A_2 = B_2-1)+P(B_2 = A_1-1)] \\ \nonumber &+&  \frac{1}{2}\big(P(A_1 = B_1 + 1) +P(B_1 = A_2+2)\\ \nonumber &+& P(A_2 = B_2 + 1) +P(B_2 = A_1 + 1 ) \\ \nonumber &-& [P(A_1 = B_1-2) P(B_1 = A_2 - 1)\\ \nonumber &+&  P(A_2 = B_2-2) +P(B_2 = A_1-2)]\big)\leq 2 
\end{eqnarray}
% and the shared entangled state
%\begin{eqnarray}
%\label{p5}
can be used to obtain violations of CGLMP  in  a sequential situation with 
\(|\psi_{\mathcal{M}_E}^5 \rangle =  \frac{1}{\sqrt{5}}(|00 \rangle +  |11 \rangle +  |22 \rangle +  |33 \rangle +  |44 \rangle)\).
%\end{eqnarray}
%nly two Bobs can shares non-locality.
 The optimal quantum violation of CGLMP inequality for Bob$_{1}$, Bob$_{2}$ and Bob$_{3}$ are given in Table. \ref{tab:mes} upto $d=10$.

\begin{table}[]
\begin{center}
\begin{tabular} {|p{1.5cm}|p{1.5cm}|p{1.5cm}|p{1.5cm}|}  
\hline
\multicolumn{4}{|c|}{Optimal quantum value of CGLMP inequality} \\
\hline
\centering Dimension & \centering Bob$_1$ & \centering Bob$_2$ & \ \ \ \ \  Bob$_3$ \\
\hline
\centering $3$ & \centering $2.8729$ & \centering $2.4086$ & \ \ \  $1.8380$  \\
\centering $4$ & \centering $2.8962$ & \centering $2.3963$ & \ \ \  $1.7994$  \\
\centering $5$ & \centering $2.9105$ & \centering $2.3819$ & \ \ \  $1.7650$  \\
\centering $6$ & \centering $2.9202$ & \centering $2.3699$ & \  \ \  $1.7382$  \\
\centering $7$ & \centering $2.9272$ & \centering $2.3570$ & \ \ \  $1.7122$ \\
\centering $8$ & \centering $2.9324$ & \centering $2.3458$ & \ \ \  $1.6910$  \\
\centering $9$ & \centering $2.9365$ & \centering $2.3360$ & \ \ \  $1.6722$  \\
\centering $10$ & \centering $2.9398$ & \centering $2.3274$ & \ \ \  $1.6568$  \\
\hline
\end{tabular}
\end{center}
\caption{Optimal quantum value of Bob$_{1}$, Bob$_{2}$ and Bob$_{3}$ are obtained using CGLMP inequality for maximally entangled state for $d=3$ to $d=10$ dimensions.}
\label{tab:mes}
\end{table}
From Table. \ref{tab:mes},  we can see that as the dimension increases, there is an increment of optimal quantum value for Bob$_1$ although the same decreases with the increase of dimension for Bob$_2$ and Bob$_3$. 
%Hence, the hope of getting more than two Bobs sharing nonlocality with Alice using CGLMP inequality for $\mathcal{M}_E$ is not possible even if the violation of CGLMP inequality increases with dimension $d$. 
 It also indicates that the trade-off between the information gained by the measurement and the disturbance created by the measurement plays a crucial role in this enterprise.

%we have studied the sharing of non-locality using CGLMP inequality for \mathcal{M}_V.

\begin{table}[]
\begin{center}
\begin{tabular} {|p{1.5cm}|p{1.5cm}|p{1.5cm}|p{1.5cm}|}  
\hline
\multicolumn{4}{|c|}{Optimal quantum value of CGLMP inequality} \\
\hline
\centering Dimension & \centering Bob$_1$ & \centering Bob$_2$ & \ \ \ \ \  Bob$_3$ \\
\hline
\centering $3$ & \centering $2.9150$ & \centering $2.4402$ & \ \ \  $1.8578$  \\
\centering $4$ & \centering $2.9729$ & \centering $2.4526$ & \ \ \  $1.8307$  \\
\centering $5$ & \centering $3.0158$ & \centering $2.4564$ & \ \ \  $1.8015$  \\
\centering $6$ & \centering $3.0495$ & \centering $2.4522$ & \  \ \  $1.7702$  \\
\centering $7$ & \centering $3.0771$ & \centering $2.4418$ & \ \ \  $1.7342$ \\
\centering $8$ & \centering $3.1012$ & \centering $2.4324$ & \ \ \  $1.7041$  \\
\centering $9$ & \centering $3.1215$ & \centering $2.4231$ & \ \ \  $1.6768$  \\
\centering $10$ & \centering $3.1393$ & \centering $2.4142$ & \ \ \  $1.6517$  \\
\hline
\end{tabular}
\end{center}
\caption{\(I_d^i\), \((i=1,2,3)\), for  Bob$_{1}$, Bob$_{2}$ and Bob$_{3}$  are listed for maximally violating state ($\mathcal{M}_V$) as the initial state from $d=3$ to $d=10$.}
\label{tab:mvs}
\end{table}

Since the  CGLMP inequality gives the maximum violation for a non-maximally entangled state, let us examine if the initial shared state in a sequential scenario is $\mathcal{M}_V$ and whether the situation improves or not. 
%We repeat similar analysis for the maximally violating states and observe that 
The observation is that although the first round  of violation is more and increases faster with $d$ in comparison to the $\mathcal{M}_E$, 
the measurements disturb the state to such an extent that violation for more than two Bobs still remains an impossibility.
Also, note that the third round value of the CGLPM expression decreases on increasing the dimension, 
so the possibility of getting a simultaneous violation for three rounds is unlikely even if $d$ is increased beyond $10$. 
See Table. \ref{tab:mvs} for details. Comparing Tables \ref{tab:mes} and \ref{tab:mvs}, we observe that the gap between the \(I_d^3\) values obtained for $\mathcal{M}_V$ and $\mathcal{M}_E$  decreases
 with the increase of dimension. It possibly indicates that the unsharp measurements disturb the $\mathcal{M}_V$ more drastically than the $\mathcal{M}_E$ in higher dimensions. 
 
\begin{figure}[ht]
\includegraphics[width=\linewidth]{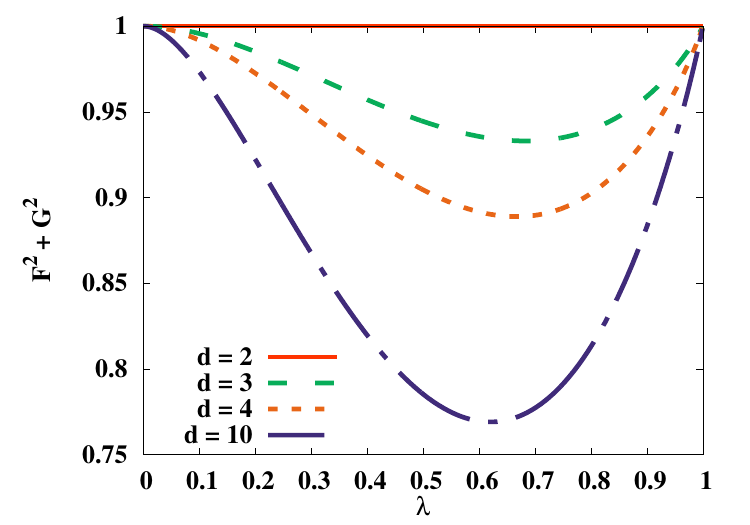}
\caption{Variation of $F^2 + G^2$ versus the sharpness parameter of measurement $\lambda$ for various $d$. As $d$ increases the the weakening strategy via white noise becomes suboptimal.}
\label{fig:op}
\end{figure}

%\textcolor{blue}{
\subsection{Optimality analysis of measurements in the sequential scenario}
An important question is whether the POVMs considered by weakening the optimal single round measurement strategy with white noise might not be the optimal one to get the maximal number of Bobs that can sequentially violate the CGLMP inequality with a single Alice. The optimality in the sequential case might be measured by the strategy that maximizes the number of Bobs that violate the CGLMP inequality. However, as one can clearly see, such a definition of optimality leaves a lot of degeneracy in the number of measurement choices that achieve the “optimal” number of Bobs, since the indicator of optimality increases in integer steps.

Ideally, one would like to perform a ``full optimization" to maximize $n$. The full optimization refers to a maximization over measurement settings of the CGLMP inequality and unsharp parameters of all the $n$ rounds collectively. Since the measurement settings of the $m$th round adaptively depend on the settings of all the previous rounds,  the full optimization process becomes realistically intractable. Even if each round is treated independently with an objective to obtain an infinitesimal amount of violation with minimal disturbance to the state, we again run into the domain of infeasibility. This is because if the optimization at every round is carried over arbitrary POVMs, the dimension of the parameter space in the optimization would become very large even if one considers POVMs with a fixed number of outcomes. This again renders the problem intractable. Therefore, we resort to optimization over a simplified setting where we are left with a five-parameter optimization at every round. Specifically, we numerically scan the measurement strategies of the form, the optimal setting used for CGLMP inequality made unsharp with  white noise, denoted by (CGLMP + white) to evaluate the optimal setting. In particular, we perform a five-dimensional optimization over $\alpha_1, \alpha_2, \beta_1, \beta_2$ and $\lambda$ to find the violation and the  post-measurement state in one round and then perform optimal projective CGLPM measurements to check whether we are getting any violation in the next round. If no, we stop, and if yes, we repeat the same drill of numerical optimization. Our analysis using DIviding RECTangles algorithm \cite{direct} for global optimization reveals that for the measurement strategies of the form (CGLMP + white), the best setting is when $\alpha_1, \alpha_2, \beta_1$, and $\beta_2$ are chosen to be the optimal setting as in the projective (single round) case of the CGLMP inequality.

Lastly we believe that our choice of weak measurements is one of the most ``intuitive" options one may consider. If the full optimization, in principle, yields some other strategy,   we suspect that this approach might constitute some exotic measurements that might be difficult to implement operationally. In such a situation, we argue that our choice of unsharp measurements is the most practically motivated one.

Now we present an analysis of the level of optimality provided by our choice of the set of weak measurements over which our optimization runs.
Given a weak measurement scheme, the optimality of such a strategy involves the study of the tradeoff between the information gain by the measurement and the amount of disturbance it imparts to the state. The amount of disturbance quantifies the quality of measurement $(F)$ and can be obtained from the post measurement state $\rho'$
\begin{eqnarray}
\rho' = F \rho + (1 - F) \sum_{i=0}^{d-1} P_i \rho P_i,
\label{eq:Fdef}
\end{eqnarray}
where $\rho$ is the initial state and $P_i$s are the projectors that have been weakened by the strategy 
\begin{eqnarray}
E_i = \lambda P_i + \frac{1-\lambda}{d} \mathbb{I}_d,
\label{eq:weakstrategy}
\end{eqnarray}
with $\mathbb{I}$ being the $d$-dimensional identity, the same strategy we have used in our work, see Eq. \eqref{pv}. Again, from Eq. \eqref{pv}, we can make the following identification, $P_i = |l = i\rangle\langle l = i|$. The second quantity of interest measures the amount of information gained in the experiment, which can be interpreted as the precision of the experiment is given by $G$, where
\begin{eqnarray}
p(E_i) = G ~tr(P_i \rho) + \frac{1-G}{d},
\label{eq:defG}
\end{eqnarray}
where $p(E_i)$ is the clicking probability of $E_i$. With $F$ and $G$, we have the following information gain vs. disturbance inequality \cite{silva'15}
\begin{eqnarray}
F^2 + G^2 \leq 1.
\end{eqnarray}
The optimal measurement strategy saturates the above inequality. In the $d = 2$ case, it was proven that weakening the optimal projective measurements via white noise can saturate the $F^2 + G^2 \leq 1$ inequality, thereby demonstrating its optimality. In the absence of the solution for the full optimization, we continue the weakening strategy via white noise in accordance with the optimal setting for $d = 2$ to higher dimensions. To test the optimality of our ansatz, for $d > 2$, we test how close to unity does $F^2 + G^2$ goes for our choice of weak measurements. For a general $d$, we compute
\begin{eqnarray}
F = \frac{2}{d}\sqrt{\big(1 + (d-1)\lambda\big)(1-\lambda)} + \frac{(d-2)(1-\lambda)}{d}.
\end{eqnarray}
See Appendix \ref{app:2} for a detailed calculation.
Furthermore, from Eq. \eqref{eq:defG}, we get  $G = \lambda$. We find that for $d>2$, $F^2 + G^2 < 1$, thereby suggesting that our measurement choices are not optimal. However, to our advantage, our analysis reveals that for low d, the inequality reaches near saturation.
%our measurement choice is near optimal. 
For example, for d = 3 and 4, we get almost $90\%$ saturation of the inequality for the relevant choice of system parameters, see Fig. \ref{fig:op}.
% We have shown that for low d, our choice of measurements turns out to be near-optimal. 

%\textcolor{red}{However we want to point out that such a definition of optimality leaves a lot of degeneracy in the number of measurement choices that achieve the “optimal” number of Bobs, since the indicator of optimality increases in integer steps. Furthermore, we want to mention that the full general optimization would involve a multi-dimensional constrained optimization which is “practically” impossible to do. So we scan a smaller set of measurement strategies where the numerical optimization can be performed in a reasonable time. We choose the set of the form (CGLMP + white) where for each round, a five-parameter optimization over $\alpha_1, \alpha_2, \beta_1, \beta_2$ and the weakness parameter $\lambda$ have to be performed.}

%\textcolor{blue}{Lastly, with the “intuitive” measurement strategies ruled out by our search, we suspect that the theoretical optimal strategy might constitute some exotic measurement that might be difficult to implement operationally. In such a situation we would like to argue that our choice of weak measurements is the most practically motivated choice that one could consider. Therefore, although we are not able to solve the full optimization problem which is quite hard and is beyond the scope of our work, we have provided reasonings and justification for the choice of measurements employed in our manuscript.}

\section{Robustness in sequential exhibition of nonlocality}
\label{sec:seqnoise}

%\section{Robustness in multiround CGLMP violation}
%\label{sec:robust2}

In Sec. \ref{sec:robust1},  we analyzed how much noise we could add to the state as well as measurements so that it continues to violate the CGLMP inequality. However, the option of using sequential measurements to obtain violations for multiple Bobs with a single Alice opens up a possibility to examine robustness from a new point of view. 
In this context, we define robustness as the maximal amount of noise that can be added to a state such that the CGLMP inequality can be violated for multiple rounds, which we claim to be two, since from the previous section, we observed that both for the $\mathcal{M}_E$ and $\mathcal{M}_V$, the maximum number of Bobs that can violate the CGLPMP inequality with Alice remains two. 

\begin{figure}[ht]
\includegraphics[width=\linewidth]{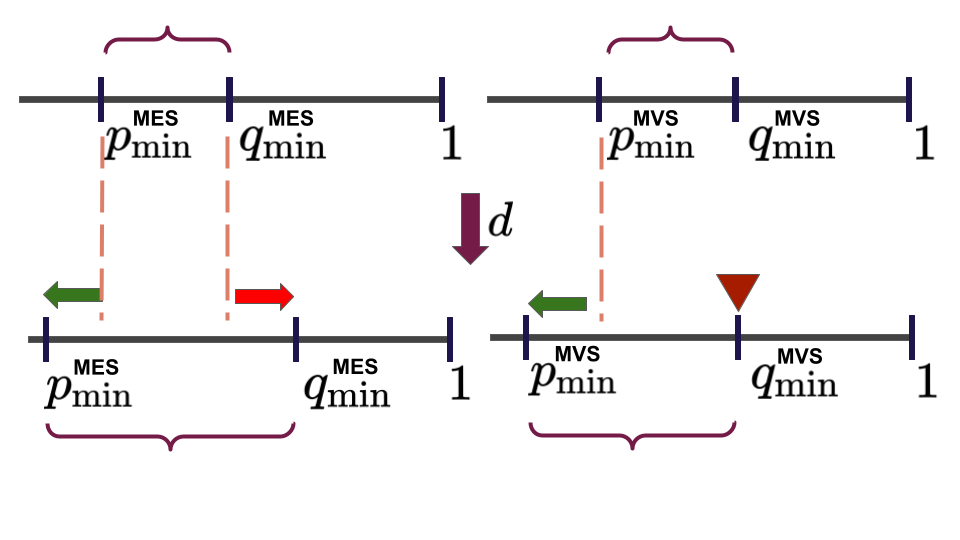}
\caption{Schematic depiction of the dynamics of $q_{\min}$ and $p_{\min}$  for both $\mathcal{M}_E$ and $\mathcal{M}_V$ with $d$. \(p_{\min}\) denotes the visibility of the state while \(q_{\min}\) is the minimum value of the visibility above which  CGLMP inequality in the second round starts violating. The superscripts represent the states considered. The green and red arrows respectively indicate the advantages and disadvantages of robustness with dimensions.  }
\label{fig:robust2}
\end{figure}

%For an initial state $|\psi\rangle$ for which two Bobs can violate the CGLMP inequality with a single Alice, 

Let us consider the pure state, $|\psi\rangle$ admixed with white noise, given in Eq. (\ref{eq:noisystate}), having the visibility, $q$, as an initial state in the sequential scenario.  We now demand that if two Bobs have to show a violation of local realism with Alice,  
%\begin{eqnarray}
%\rho_d = q|\psi\rangle\langle\psi| + \frac{(1-q)}{d^2} \mathbb{I}_{d^2},
%\end{eqnarray}
%such that
 both $I_d^1$ and $I_d^2$ have to be greater than $2$. 
 %  It indicates violation by dual Bobs. 
 We define $q_{min}$ to be the minimum value of $q$ above which both $I_d^1 > 2$ and $I_d^2 > 2$. We now compute how the $q_{\min}$ scales with $d$ and compare it with the scaling obtained for $p_{\min}$ as discussed in Sec. \ref{sec:robust1} for both $\mathcal{M}_E$ and $\mathcal{M}_V$.
\begin{table}[]
\begin{center}
\begin{tabular} {|p{1.5cm}|p{1.5cm}|p{1.5cm}|}  
\hline
\centering Dimension & \centering $q_{\min}^{\text{$\mathcal{M}_V$}}$ & \hspace{0.25cm} $q_{\min}^{\text{$\mathcal{M}_E$}}$ \\
\hline
\centering $3$ & \centering $0.8773$ & \hspace{0.225cm}  $0.8845$  \\
\centering $4$ & \centering $0.8748$ &\hspace{0.225cm} $0.8872$  \\
\centering $5$ & \centering $0.8737$ & \hspace{0.225cm} $0.8900$   \\
\centering $6$ & \centering $0.8736$ & \hspace{0.225cm} $0.8933$   \\
\centering $7$ & \centering $0.8738$ & \hspace{0.225cm} $0.8963$  \\
\centering $8$ & \centering $0.8741$ & \hspace{0.225cm} $0.8987$   \\
\centering $9$ & \centering $0.8748$ & \hspace{0.225cm} $0.9012$   \\
\centering $10$ & \centering $0.8752$ &\hspace{0.225cm} $0.9034$   \\
\hline
\end{tabular}
\end{center}
\caption{The $q_{\min}$ values for maximally entangled states ($\mathcal{M}_E$) and maximally violating states ($\mathcal{M}_V$) for $d = 3$ to $10$ are reported when we demand violation of CGLMP inequality by two Bobs sequentially with Alice.}
\label{tab:r2}
\end{table}

Recall that in CGLMP test, we observed an enhanced amount of robustness (as defined in terms of persistence of the violation on the addition of white noise) on increasing $d$ as indicated by lowered values of $p_{\min}$. The maximal amount of white noise that the state can absorb such that the violation persists is simply given by $1- p_{\min}$. %\textcolor{red}{Report properties.}
For both $\mathcal{M}_E$ and $\mathcal{M}_V$, $p_{\min}$ decreases with $d$ \cite{kas'00, cglmp'02, acin'02}.
Furthermore, note that we expectedly find  $p_{\min} < q_{\min} < 1$. 

When robustness is analyzed in the context of sustaining dual-round violation via the use of sequential measurements, we observe a qualitatively different trend. For $\mathcal{M}_E$,  $q_{\min}$  actually increases with $d$. This implies that robustness actually decreases with $d$ when $\mathcal{M}_E$ are employed and we demand CGLMP violations by two Bobs.
%when viewed from the perspective of multiple round violation.
 However, for $\mathcal{M}_V$, $q_{\min}$ values do not change significantly on increasing $d$. See Table. \ref{tab:r2} for details of the $q_{\min}$ values for both $\mathcal{M}_E$ and $\mathcal{M}_V$. However, in both cases, the gap between $q_{\min}$ and $p_{\min}$ increases with $d$. For a pictorial representation of the situation, see Fig. \ref{fig:robust2}.

The above results explain in part why despite an increase in the first round violation with $d$, one does not get a higher number of Bobs which sequentially violates CGLMP inequality i.e., $ I_d^k >2$ with $k>2$ for higher dimensional systems. Although the amount of maximal first-round violation grows, the disturbance induced by the measurements is high enough to actually bring down the violation in the second round with $d$ which ultimately leads to the third round becoming non-violating. 

\section{Discussion}
\label{sec_disc}
To achieve quantum advantage, manipulating and analysing higher dimensional quantum systems is essential, since, in several quantum information processing tasks, higher dimensional quantum systems turn out to be more beneficial than the qubit pairs. CGLMP inequality is a family of tight Bell inequalities for bipartite systems of arbitrary dimension, which is known to exhibit more robustness against noise with increasing dimension. Therefore, it is interesting to investigate how CGLMP inequality responds if noise is present not only in the state but also in measurement.

 We introduced a new measure of robustness which we referred to as the `area of nonlocal region' under consideration of dual noises, both in states and measurements. In particular, this area indicates the region in noise parameter space where violation of CGLMP can be observed. We found that this region grows more rapidly with the increase of dimension with respect to the increase in visibility associated solely with states or measurements. 
 
The introduction of noise in measurement facilitates to invoke of a sequential violation of CGLMP inequality as it retains some residual correlation after obtaining a violation in the first round. %\textcolor{red}{It addresses a fundamental question of whether with respect to the CGLMP inequality, non-locality can be shared between multiple observers. From a practical point of view, such analysis can be useful in situations where there is a significant constraint in the preparation of entangled states, for example in the Nitrogen Vacancy Centre experiments.} 
Interestingly, we found that the violation of CGLMP inequality by two sequential observers on one side and another observer on the other end persists with dimension. Moreover, the minimum visibility required to achieve double violation in the sequential case increases with the increase of dimension, thereby exhibiting the opposite behaviour as compared to the violation obtained for the shared state without unsharp measurement. It indicates that robustness in the sequential measurement scenario is qualitatively distinct from that of the typical Bell test since it involves the disturbance of the state introduced via the measurement. It is interesting to probe further how the double violation obtained in CGLMP inequality enables applications in the context of information processing tasks involving higher dimensional quantum systems.

\section*{Acknowledgement}
%We acknowledge the support from Interdisciplinary Cyber Physical Systems (ICPS) program of the Department of Science and Technology (DST), India, Grant No.: DST/ICPS/QuST/Theme- 1/2019/23. This research was supported in part
%by the INFOSYS scholarship for senior students.

We acknowledge the support of the Interdisciplinary
Cyber Physical Systems (ICPS) program of the Department of Science and Technology (DST), India, Grant No.
DST/ICPS/QuST/Theme- 1/2019/23. This research was
supported in part by the INFOSYS scholarship for senior students. S.R. acknowledges support by the Hong Kong Research
Grant Council through Grants No. 17307520 and No. R703521F, by the Chinese Ministry of Science and Technology
through Grant No. 2023ZD0300600, and by the John Templeton Foundation through Grant No. 62312, The Quantum Information Structure of Spacetime (qiss.fr). S.M. acknowledges Ministry of Science and Technology in Taiwan (Grant
No. 110-2811-M-006-501). A.K. acknowledges the research associateship from S.N. Bose National Centre of Basic Sciences, Kolkata, India.

\appendix
\section{The maximally violating states}
\label{app:1}
The expressions for the maximally violating states are given in \cite{Fonseca'18}. We state them here once more for the convenience of the readers.
\begin{widetext}
\begin{eqnarray}
	\ket{\psi_{\rm{\mathcal{M}_V}}^{d=3}}&=&0.6169\ket{00}+0.4888\ket{11}+0.6169\ket{22},\\
	\ket{\psi_{\rm{\mathcal{M}_V}}^{d=4}}&=&0.5686\ket{00}+0.4204\ket{11}+0.4204\ket{22}+0.5686\ket{33},\\
	\ket{\psi_{\rm{\mathcal{M}_V}}^{d=5}}&=&0.5368\ket{00}+0.3859\ket{11}+0.3548\ket{22}+0.3859\ket{33}+0.5368\ket{44},\\
	\ket{\psi_{\rm{\mathcal{M}_V}}^{d=6}}&=&0.5137\ket{00}+0.3644\ket{11}+0.3214\ket{22}+0.3214\ket{33}+0.3644\ket{44}+0.5137\ket{55},\\
	\ket{\psi_{\rm{\mathcal{M}_V}}^{d=7}}&=&0.4957\ket{00}+0.3493\ket{11}+0.3011\ket{22}+0.2882\ket{33}+0.3011\ket{44}+0.3493\ket{55}+0.4957\ket{66}, \\
	\ket{\psi_{\rm{\mathcal{M}_V}}^{d=8}}&=&0.4812\ket{00}+0.3379\ket{11}+0.2872\ket{22}+0.2679\ket{33}+0.2679\ket{44}+0.2872\ket{55}+0.3379\ket{66}  \nonumber \\
	&+& 0.4812\ket{77}, \\
	\ket{\psi_{\rm{\mathcal{M}_V}}^{d=9}}&=&0.4690\ket{00}+0.3288\ket{11}+0.2770\ket{22}+0.2541\ket{33}+0.2474\ket{44}+0.2541\ket{55}+0.2770\ket{66} \nonumber \\
	&+&  0.3288 \ket{77} + 0.4690 \ket{88}, \\
	\ket{\psi_{\rm{\mathcal{M}_V}}^{d=10}}&=&0.4587\ket{00}+0.3212\ket{11}+0.2690\ket{22}+0.2440\ket{33}+0.2334\ket{44}+0.2334\ket{55}+0.2440\ket{66} \nonumber \\
	&+& 0.2690 \ket{77} + 0.3212 \ket{88} + 0.4587 \ket{99}.
	\end{eqnarray}
\end{widetext}
\section{Computation of measurement quality index $F$}
\label{app:2}
Following the weak measurement strategy given in Eq. \eqref{eq:weakstrategy}, the post measurement state reads as 
\begin{eqnarray}
\rho' = \sum_{i = 0}^{d-1} M_i \rho M_i^\dagger,
\end{eqnarray}
where $M_i$s are the update operators given by
\begin{eqnarray}
M_i = \sqrt{E_i} = x P_i + y (\mathbb{I} - P_i)
\end{eqnarray}
with
\begin{eqnarray}
x = \sqrt{\frac{1 + (d-1)\lambda}{d}}, ~y = \sqrt{\frac{1-\lambda}{d}}.
\end{eqnarray} 
See Eq. \eqref{eq:weakstrategy} for the form of $E_i$s in terms of $P_i$s. Now $\rho'$ can be written as
\small
\begin{eqnarray}
\rho' &=& \sum_{i=0}^{d-1} P_i \rho P_i + (2xy + (d-2)y^2) \sum_{\substack{ij=0 \\ i \neq j}}^{d-1} P_i \rho P_j, \nonumber \\
&=& (2xy + (d-2)y^2) \rho + (1- 2xy - (d-2)y^2) \sum_{i=0}^{d-1} P_i \rho P_i. \nonumber \\
\end{eqnarray}  
\normalsize
Now following Eq. \eqref{eq:Fdef}, we make the following identification for the quality factor of the measurement
\begin{eqnarray}
F &=& 2xy + (d-2)y^2, \nonumber \\
&=& \frac{2}{d}\sqrt{\big(1 + (d-1)\lambda\big)(1-\lambda)} + \frac{(d-2)(1-\lambda)}{d}.
\end{eqnarray}

\end{document}